\documentclass[aps,prl,twocolumn,superscriptaddress]{revtex4-2}
\usepackage{amsmath, amssymb}
\usepackage{bbold}
\usepackage{graphicx}
\usepackage{float}
\usepackage{natbib}
\usepackage[colorlinks=true,linkcolor=blue,urlcolor=black,citecolor=blue]{hyperref}

\usepackage{pdfpages}
\usepackage{pgffor}
\makeatletter
\AtBeginDocument{\let\LS@rot\@undefined}
\makeatother

\newcommand{\prltshs}[1]{{\normalfont\itshape{#1}.\textbf{--}}}

\makeatletter
\renewcommand\section{\@startsection {section}{1}{1em}%section, level 1, indent 1em
                                   {-1ex}%to prev. line space, -1=no
                                   {-1ex}%...next...
                                   {\prltshs}}%style
\makeatother
\newcommand{\crb}{Cram\'er-Rao bound}

\begin{document}

\title{Learning Non-Markovian Noise via Ensemble Optimal Control}

\author{Da-Wei Luo}
\email{dawei.luo@stevens.edu}
\affiliation{Center for Quantum Science and Engineering and Department of Physics, Stevens Institute of Technology, Hoboken, New Jersey 07030, USA}

\author{Ting Yu}
\email{Ting.Yu@stevens.edu}
\affiliation{Center for Quantum Science and Engineering and Department of Physics, Stevens Institute of Technology, Hoboken, New Jersey 07030, USA}

\date{\today}

\begin{abstract}

We study the estimation of parameters pertaining to non-Markovian quantum open systems, such as the dissipation rate and environmental memory time. A key challenge is identifying the optimal measurement time, which must allow sufficient time to acquire information about the environment, yet be short enough to avoid dissipation that erases the information.

Using machine learning approaches, we develop an optimized control scheme trained over a representative ensemble to fix the optimal measurement time at a prescribed runtime. The protocol is robust to errors in the training process, enhances precision by exploiting non-Markovian memory effects, and achieves measurement uncertainties approaching the quantum limits set by the Cram\'{e}r-Rao bound.

\end{abstract}
\maketitle

%%%%%%%%%%%%%%%%%%%%%%
\section{Introduction}

Advances in quantum sensing and metrology have enabled sensitive and precise probing of physical systems, with applications ranging from weak force and field sensing~\cite{Degen2017a,Korobko2023a,Weiss2021a} and gravitational wave detection~\cite{Abramovici1992a,Goda2008a} to biomedical technologies~\cite{Aslam2023a,Zheng2024a}. By exploiting quantum resources such as coherence, entanglement, and squeezing~\cite{Huelga1997a,Agarwal2022a,Degen2017a,DeMille2024a,Demkowicz-Dobrzaifmmode-nelse-nfiski2014a,Kamble2024a,Korobko2023a,Linnemann2016a,Wang2024a,Xia2023a,Yang2024a}, these platforms can surpass classical precision limits.

In realistic implementations, however, coupling to the environment introduces noise and open-system effects~\cite{Breuer2002,Breuer2016a,Shrikant2023a,Vega2017a}. In particular, non-Markovian (colored) noise with memory can significantly influence system dynamics~\cite{Breuer2016a}, enabling enhanced controllability~\cite{Bhaktavatsala-Rao2011a,Kofman2001a,Floether2012a,Li2020a} and improved state transfer and entanglement generation~\cite{Mu2016a,Yu2023a,Zhao2011a}. Accurate characterization of such noise is therefore essential. It has also been shown that for non-Hermitian Hamiltonians~\cite{Arkhipov2026a,Bao2021a,Bao2024a}, non-reciprocity can be a powerful resource in quantum sensing, and may be utilized to construct exponentially sensitive sensors.

Unlike closed systems where longer evolution typically improves parameter estimation precision~\cite{Garbe2022a,Lecamwasam2024a}, quantum open systems exhibit a trade-off between information gain and dissipation~\cite{Saleem2023a,Saleem2024a}. The system must evolve long enough to encode information about the environment, yet not so long that this information is degraded. This leads to the central challenge of identifying an optimal measurement time, which itself may depend on the unknown parameters.

In this Letter, we propose a protocol to address the challenge of unknown optimal measurement time by training an ensemble-based optimal control~\cite{Goerz2014a,Goerz2019a}. Rather than determining the optimal time explicitly, we employ an active, open-loop control to enforce optimal measurement at a prescribed runtime. The control pulse is obtained using a machine-learning-inspired approach, where it is trained over a representative set of parameters and subsequently validated both within and beyond the training regime.

Using numerical automatic-differentiation techniques~\cite{Goerz2022a,Innes2019a}, we directly optimize the measurement precision at a fixed time by maximizing the quantum Fisher information (QFI)~\cite{Qin2022a,Helstrom1969a,Petz2011a}. As illustrative examples, we consider the estimation of non-Markovian noise parameters, including memory time and coupling strength. Although the training assumes a parameter range and fixed system settings, the resulting control is robust to deviations in both, and effectively drives the system toward states optimal for parameter estimation.

Notably, non-Markovian memory effects enhance the achievable precision. The resulting measurement uncertainty approaches the limits set by the the Cram\'{e}r-Rao bound.~\cite{Cramer1999a,Petz2011a,Helstrom1969a}, while remaining monotonic-and thus invertible-allowing direct parameter extraction via simple curve fitting. We first present the general protocol and then demonstrate its performance for a two-level system. The approach is broadly applicable to systems in which the QFI does not increase monotonically with time.

%%%%%%%%%%%%%%%%%%%%%%
\section{Learning open system parameters using Krotov ensemble optimization control}

Consider a quantum system embedded in a bosonic bath, described by the total Hamiltonian~\cite{Breuer2016a}
\begin{align}
	H_{\rm tot} = H_s + \sum_k \omega_k b_k^\dagger b_k + \sum_k \left( g_k^* L b_k^\dagger + g_k L^\dagger b_k \right),
\end{align}
where $H_s$ is the system Hamiltonian, $b_k$ is the annihilation operator for the $k$-th bath mode with frequency $\omega_k$, $L$ is the system-bath coupling operator, and $g_k$ denotes the coupling strength between the system and the $k$-th bath mode. Tracing out the bath degrees of freedom then gives the reduced non-unitary system dynamics, which can be specified as a master equation or an unraveled quantum trajectory form. Notably, the influence of the bath on the system may be summarized by a bath spectrum $J(\omega) = \sum_k |g_k|^2 \delta(\omega-\omega_k)$, or via its Fourier transform, known as the bath correlation function $\alpha(t,s)=\int d\omega J(\omega)e^{-i \omega(t-s)}$, assuming a zero-temperature bath. Finite-temperature baths may be cast as a fictitious thermal-vacuum state~\cite{Yu2004a}; thus, we will focus on the zero-temperature bath for simplicity here.

In the limit of weak system-bath coupling and a forgetful bath whose memory effects are ignored, one would have a Markov open system whose spectrum $J(\omega)$ is flat, and the only parameter of interest is the system-bath coupling strength, akin to white noise. However, when no such approximations are made and non-Markovian memory effects are taken into full consideration, the non-flat bath spectrum can exhibit interesting features and represents a more faithful description of the open system dynamics. One hallmark feature of non-Markovian dynamics is a backflow of information from the bath into the system, which can have subtle influences on the system dynamics, such as facilitating entanglement generation~\cite{Mu2016a,Zhao2011a} or offering better control of the system dynamics~\cite{Bhaktavatsala-Rao2011a,Kofman2001a,Floether2012a,Li2020a,Yu2023a} and sensing~\cite{Altherr2021a,Chin2012a}. One commonly used spectrum is the Lorentzian spectrum,
\begin{align}
	J(\omega) = \frac{1}{2\pi}\frac{\Gamma \gamma^2}{(\omega-\Omega)^2 + \gamma^2},
\end{align}
with a corresponding bath correlation function $\alpha(t,s)=\Gamma \gamma \exp[-\gamma|t-s|-i \Omega(t-s)]/2$, where $\Gamma$ denotes the system-bath coupling strength, $\gamma$ is the memory parameter, and $\Omega$ represents a central frequency shift. The Lorentzian spectrum has the interesting feature that smaller $\gamma$ dictates a stronger memory effect, and in the limit $\gamma \rightarrow \infty$, it continuously approaches the Markov limit, which enables one to probe bath memory effects with the $\gamma$ parameter. Instead of studying parameter estimation through an abstract map~\cite{Kamble2024a,Wang2024a}, here we consider the problem as a dynamical one. Sensing of bath properties has been discussed for specific systems using encoding techniques~\cite{Tan2022a} or with special initial states that utilize bound states of the open system~\cite{Wu2021a}. As an alternative approach, here we discuss the use of quantum control for bath sensing, which could be more general and applicable to a wider class of open systems. Also note that while the use of control in quantum sensing has been studied in the context of non-Markovian dynamics~\cite{Yang2024b,Qin2022a}, it is mostly concerned with the estimation of system parameters rather than noise parameters, and the problem of needing to know the parameter in order to derive the control remains.

With applications in quantum metrology in mind, we consider the problem of measuring the bath spectrum parameters $\Gamma$, $\gamma$, and $\Omega$ as a dynamical process. The sensitivity of the measurement protocol can be quantified by the quantum Fisher information (QFI)~\cite{Helstrom1969a,Petz2011a,Liu2019a} $F_\theta(t) = \mathrm{tr}(\rho(t) L_\theta^2)$, where $L_\theta$ is the symmetric logarithmic derivative (SLD) operator for a parameter $\theta$, defined as $\partial_\theta \rho = \left( \rho L_\theta + L_\theta \rho \right)/2$. At $t=0$, one would have no knowledge of the open system parameters, $F_\theta(0)=0$, while at some asymptotically large time $T_\infty$, under open system dynamics the state is most likely to relax into a steady state and the knowledge of the open system parameters would be lost, $F_\theta(T_\infty)\sim 0$. It is easy to see that  the Fisher information would then have at least one peak between the two time points, as long as it is not identically zero.

\begin{figure}[t]
    \centering
    \includegraphics[width=.4\textwidth]{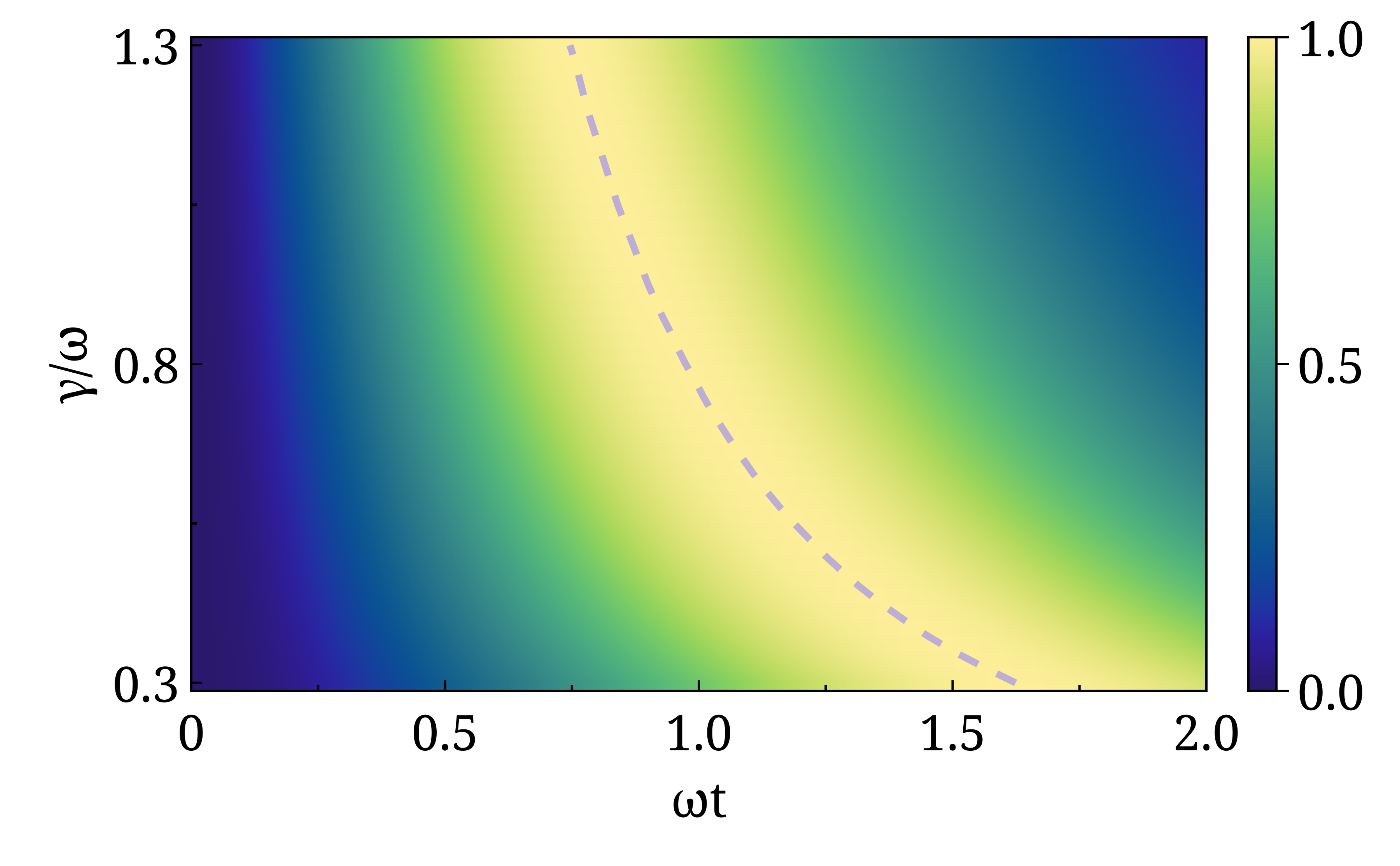}
    \caption{(Color online) For an example of the TLS dephasing model with $\Gamma=1$, $\Omega=0$ and under different memory parameters between $\gamma=0.3$ and $\gamma=1.3$. The normalized quantum Fisher information is plotted as a function of time $t$ and the memory parameter $\gamma$, scaled by their respective maximum values, and the dashed line signifies the time it takes for the QFI to reach its peak. Without knowing the parameters beforehand, it is impossible to know the optimal time to take the measurement, which can be quite far apart.}\label{fig_2pks}
\end{figure}

The operational issue in measuring the parameters is then the determination of an optimal time to take the measurement, such that the system has evolved long enough to accumulate sufficient information on the open system parameters, but not so long that the obtained information is lost. In terms of the Fisher information, we want to make the measurement when $F$ is at or near its maximum value. However, without knowing the parameter beforehand, it is impossible to determine when to take the measurement. For example, with a two-level system dephasing model~\cite{Diosi1998a} $H_s=L=\sigma_z$, we plot the normalized QFI as a function of time between two different memory parameters, $\gamma=0.3$ and $\gamma=1.3$, in Fig.~\ref{fig_2pks}. The optimal measurement time for them is quite different, and finding an ideal time that works across an estimated parameter range can be suboptimal.

As a solution to this quantum sensing task, we propose a protocol based on optimal control and machine learning. The first part of the protocol relies on quantum optimal control; specifically, we utilize Krotov control~\cite{Konnov1999b,Reich2012v,Tannor1992h}, which is applicable in general to linear systems of the form
\begin{align}
\frac{\partial s_k(t)}{\partial t} = \left[M_0 + \sum_{i=1} c_i(t) M_i\right] s_k(t), \label{eqn_kc_dsdt}
\end{align}
where $s_k(t)$ is a state vector labeled by $k$, $M_0$ is the uncontrolled part of the system (which can be time-dependent), and $M_i$ are time-independent matrices denoting the controlled part for $i=1...n_c$ with $n_c$ control fields. The Schr\"odinger equation can be written in this general form with $M_i \sim -iH_i$ and $s_k=|\psi_k \rangle$, and the vectorized master equation $\partial_t |\rho \rangle \rangle = \mathcal{L}|\rho \rangle \rangle$ for open-system dynamics can also be written in this linear form~\cite{Goerz2019a,supp}.

The Krotov method guarantees monotonic convergence for the minimization of an optimization functional of the form
\begin{align}
J_i\left[s_k(t), c_l(t)\right] = J_T[s_k(T_f)] + \sum_l \int_0^{T_f} g_a[c_l(t)] dt, \label{eqn_kc_jt}
\end{align}
for the $i$-th iteration, where $s_k(t)$ represents the quantum state labeled by $k$ at time $t$, $c_l(t)$ is the $l$-th control function, $g_a$ is a running cost of the controls, and $J_T$ is the main part of the control functional to be minimized. The Krotov method takes a guess control as a reference and updates it for the next iteration such that monotonic convergence is guaranteed, $\Delta J_{i} < 0$, up to the effects of time discretization.

Here, if we choose the main target functional $J_T = -F_\theta(s)$ for $\theta \in {\gamma, \Gamma, \Omega}$~\cite{supp,Qin2022a,Liu2019a}, we may directly maximize the QFI with respect to (wrt.) the open-system parameters \textit{at a prescribed time}. Note that since the QFI depends not only on the quantum state $\rho(T_f)$ but also on its derivative wrt the parameter under consideration, we set $s$ to be the concatenated vectorized density operators with finite difference $dx$ as a direct sum, $s_k(t) = |\rho_\theta(t) \rangle \oplus |\rho_{\theta+dx}(t) \rangle$~\cite{supp}.

In addition, the Krotov method requires the calculation of the derivative of the target function with respect to the evolved state as a co-state, $\chi=-\partial J_T/\partial s_k(T_f)$. While analytically this is a hard problem and some workarounds have been proposed to optimize the Hilbert-Schmidt distance instead of the QFI, it has been noted that such approaches do not guarantee improvement of the QFI~\cite{Qin2022a}. To systematically address this issue, the use of numerically exact automatic-differentiation (AD) techniques~\cite{Goerz2022a,Innes2019a,supp} has been proposed to calculate the co-state derivative, allowing one to use more complex functions as control targets that are otherwise analytically intractable. With this approach, we may now directly use the QFI as the optimization functional.

%%%%%%%%%%%%%%%%%%%

On its own, the Krotov control does not fully solve the problem since, to derive the control function, we still need to know the exact value of the parameters to be estimated. For the second part of our protocol, we first note that the Krotov control allows for \textit{different} effective Hamiltonians (or Lindbladians) in the form of an ``ensemble optimization''~\cite{Goerz2014a,Goerz2019a}, where one samples representative configurations and optimizes over an average of the ensemble. Specifically, we can have \textit{different} $M_i$ in Eq.~\eqref{eqn_kc_dsdt} for each state $s_k$, while referencing \textit{the same} control fields $c_i(t)$:
\begin{align}
\frac{\partial s_1(t)}{\partial t} &= \left[M_0^{(1)} + \sum_{i=1} c_i(t) M_i\right] s_1(t), \nonumber \\
\frac{\partial s_2(t)}{\partial t} &= \left[M_0^{(2)} + \sum_{i=1} c_i(t) M_i\right] s_2(t), \ldots
\end{align}
and the new optimization functional can be a simple average $\bar{J}\left[s_k(t), c_l(t)\right]$ over $s_k(t)$. Here, by choosing different $M_0^{(j)}$ as the effective non-Markovian Lindbladians under different open-system parameters, $M_0^{(j)} = \mathcal{L}_0(\theta_j) \oplus \mathcal{L}_0(\theta_j+dx)$, where $\theta_j \in {\gamma, \Gamma, \Omega}$ is the parameter to be estimated, we are now equipped with a way to find a control field that maximizes the average QFI at a prescribed time $T_f$, for representative choices of the bath parameters $\theta \in R$.

We can then borrow an idea from machine learning, where we ``train'' a control that optimizes the QFI for a parameter assumed to lie in some training range $R=[a,b]$ by using ensemble optimization over $n$ points in $R$. We may then verify the validity of the control function by testing a large number of random parameters both inside and outside $R$. Thus, this protocol allows one to attain an optimal QFI at a prescribed time even with limited knowledge of the system parameters to be estimated: the only assumption is that the unknown parameter lies in or near some range $R$.

As an example, we consider the two-level system (TLS) dephasing model $H_{0}=L=\sigma_z$, where $\sigma_{x,y,z}$ are the Pauli matrices, and the control Hamiltonian is chosen to be $H_c = (\sigma_x+\sigma_y)/2$, with the total system Hamiltonian given by $H_s(t) = \omega H_0 + c(t)H_c$. The non-Markovian dynamics of the system may be solved by the Quantum State Diffusion (QSD) approach~\cite{Diosi1998a,qsd_n1,Yu1999a}, which projects the bath modes onto Bargmann~\cite{Vourdas1994a} coherent states labeled as $|z^*_k\rangle$, such that the open-system effects can be encapsulated by a noise $z_t^*=-i\sum_k g_k z^*_k e^{i \omega_k t}$, which, under the Lorentzian spectrum considered, is an Ornstein-Uhlenbeck process.

The quantum states are then governed by a stochastic Schr\"odinger equation $\partial_t |\psi_t(z^*) \rangle = \left[-i H_{s}(t) + Lz^*_t - L^\dagger \bar{O}(t)\right] |\psi_t(z^*) \rangle$, where the $\bar{O}$ operator represents the functional derivative of the stochastic trajectories with respect to the noise, $O(t,s)=\delta |\psi_t(z^*) \rangle/\delta z_s^*$, and $\bar{O}(t)=\int_0^t ds \alpha(t,s)O(t,s)$. While generally noise-dependent, the leading order of the $\bar{O}$ operator may be derived~\cite{Diosi1998a,Yu1999a},
\begin{align}
\partial_t O(t,s) \approx \left[-iH_{s}(t) - L^\dagger \bar{O}(t), O(t, s)\right]. \label{eq_dodt}
\end{align}

The reduced density operator is then given as the ensemble average of the trajectories $\rho(t)=\mathcal{M}\left[ |\psi_t(z^*) \rangle \langle\psi_t(z^*) |\right]$, with $\mathcal{M}[F]=\prod_k\int \exp(-|z|^2) F  d^2z /\pi$, and a corresponding master equation can be derived using Novikov's theorem~\cite{Yu2004a,Novikov1965a},
\begin{align}
\frac{\partial}{\partial t} \rho(t) &=-i \left[H_s(t),\rho(t)\right] \nonumber \\
&+\left[L,\rho(t)\bar{O}^{\dagger}(t)\right]-\left[L^\dagger,\bar{O}(t)\rho(t)\right], \label{eq_meqn}
\end{align}
which can then be vectorized by column-stacking the density operator, denoted as $\partial_t |\rho \rangle\rangle =\mathcal{L} |\rho \rangle\rangle = \left[\mathcal{L}_0(\gamma, \Gamma, \Omega) + c(t)\mathcal{H}_c \right]| \rho \rangle\rangle$, where we single out the uncontrolled part as $\mathcal{L}_0$ and the controlled part $-i[H_c, \rho]$ as $\mathcal{H}_c$.

The control problem for the optimization of the QFI may now be written as a direct sum involving a finite difference $dx$. For example, to estimate the memory parameter $\gamma$, the dynamics of the system to be controlled can be written as
\begin{align}
s'(t) &= \left[\mathcal{L}_0(\gamma, \Gamma, \Omega) \oplus \mathcal{L}_0(\gamma+dx, \Gamma, \Omega) + c(t)\mathcal{H}_c^\oplus\right] s(t),
\end{align}
where $s(t)=|\rho_\gamma(t) \rangle \rangle \oplus |\rho_{\gamma+dx}(t) \rangle \rangle$, $\mathcal{H}_c^\oplus = \mathcal{H}_c \oplus \mathcal{H}_c$.

\begin{figure}[t]
    \centering
    \includegraphics[width=.42\textwidth]{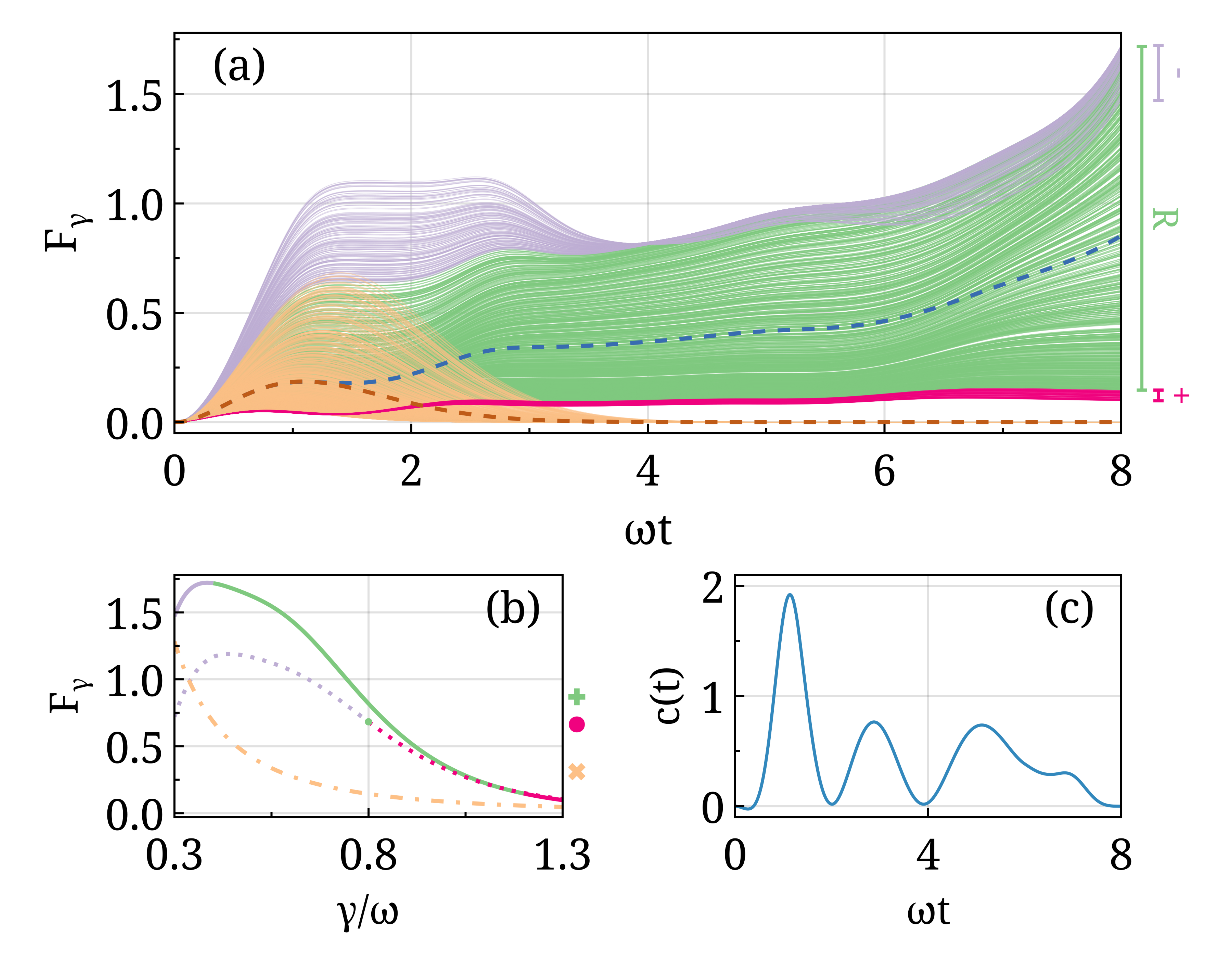}
    \caption{(Color online) Learning the memory parameter with trained control. Panel (a): Fisher information wrt. $\gamma$ as a function of time, as the uncontrolled case (orange line), $1000$ random tests: $600$ tests in the trained range (green lines), and $200$ random tests for above (below) the training range (magenta and purple lines respectively). Training data is taken from $\gamma \in [0.4, 1.2]$. The averages of the uncontrolled and trained tests are shown as dashed lines. Panel (b): Solid lines: Final Fisher information as a function of $\gamma$, color scheme as in panel (a). Dotted line shows the final QFI from optimizing just for the mid-point $\gamma_M=0.8\omega$. Dash-dotted orange line shows the peak Fisher information of uncontrolled cases $\max(F_0)$. The markers to the right shows the average values (green $+$ for the ensemble optimization, red circle for a single optimization of $\gamma_M$ and orange $\times$ for the peak of the uncontrolled case). Panel (c): Control function as a function of time.}\label{fig_sz_gamma}
\end{figure}

Note that, unlike the Markov case where the dissipative part of the master equation is independent of the system Hamiltonian, for non-Markovian dynamics the dissipative term in Eq.~\eqref{eq_meqn} is given by the $\bar{O}$ operator, which depends on the system Hamiltonian via Eq.~\eqref{eq_dodt}. The implication of this dependence is twofold: first, we need to update the $\mathcal{L}_0$ part after each iteration of the Krotov control algorithm, since the updated $c(t)$ gives a different dissipative term~\cite{supp}. Thus, the Krotov inverse step size $\lambda_a$ needs to be sufficiently large so that this reset does not introduce numerical instability and preserves monotonic convergence. Another complication is that, to calculate the Krotov control update, we approximate $\partial \mathcal{L}/\partial {c(t)}|_{t=T_f} \approx \mathcal{H}_c$, i.e., $\partial \mathcal{L}_{0}/\partial {c(t)}|_{t=T_f} \approx 0$. This can be justified analytically in the post-Markov case, and numerically it can be shown that a sufficiently large inverse Krotov step size $\lambda$ and time-discretization size $N_t = T_f/dt$ lead to convergent results~\cite{supp}.

Taking the estimation of the memory parameter $\gamma$ in a dephasing TLS as an example, we plot the main results in Fig.~\ref{fig_sz_gamma}. For training the ensemble optimization, we assume $\gamma$ to be in a training range $R = [0.4, 1.2]$. Empirically, weaker memory effects (larger $\gamma$) makes it more difficult to estimate the memory parameter so we put more sample points in this range, where a total of $60$ representative points are taken in $R$ with $20$ evenly spaced points in $[0.4, 0.7]$ and $40$ in $(0.7, 1.2]$ for the training of the ensemble optimization. The total runtime is set as $T_f=8$, with $\omega=1, \Gamma=\omega$ and $\Omega=0$, and an initial state $(|0 \rangle + |1 \rangle)/\sqrt{2}$. In panel (a), we show the dynamics of the QFI for the uncontrolled system as orange lines, and the controlled QFI in (above, below) $R$ are shown as green (magenta, purple) lines respectively. It can be seen that we can achieve maximized QFI at the prescribed time $T_f$ for parameters both in and outside the training region $R$, whereas the uncontrolled dynamics will have peak QFI at different times depending on the parameters. In Fig.~\ref{fig_sz_gamma} panel (b) we show the final QFI as a function of $\gamma$ as solid lines. The QFI is found to be larger for smaller $\gamma$ in $R$, indicating that non-Markovian memory effects here can improve the precision of the quantum sensing. The maximum QFI for the uncontrolled QFI $\mathrm{max}(F_0)$ under the same parameters is shown as the orange dash-dotted line. We may observe that not only can we achieve an optimal QFI at the end of the prescribed runtime $T_f$, the achieved QFI is also found to be larger than the uncontrolled case. It is also worth pointing out that depending on the system and parameters, while it may be possible to find a control that is relatively robust for a parameter in some range $R$ by, for example, optimizing only for the mid-point, such a technique may not always work, and the ensemble optimization control can be a more systematic and reliable way to find a control that works for parameters in an assumed range. As example, in Fig.~\ref{fig_sz_gamma}(b) we also show the final QFI for the case where a single optimization is made for the mid-point $\gamma_M=0.8\omega$. It can indeed be seen while it is possible to drive the QFI in this range, the final QFI is lower than the ensemble optimization~\cite{supp}.
We also display the control function $c(t)$ as a function of time in panel (c) of Fig.~\ref{fig_sz_gamma}, where for practical reasons it's often desirable to be well-behaved so it's neither too large nor oscillates too fast.

\begin{figure}[t]
    \centering
    \includegraphics[width=.42\textwidth]{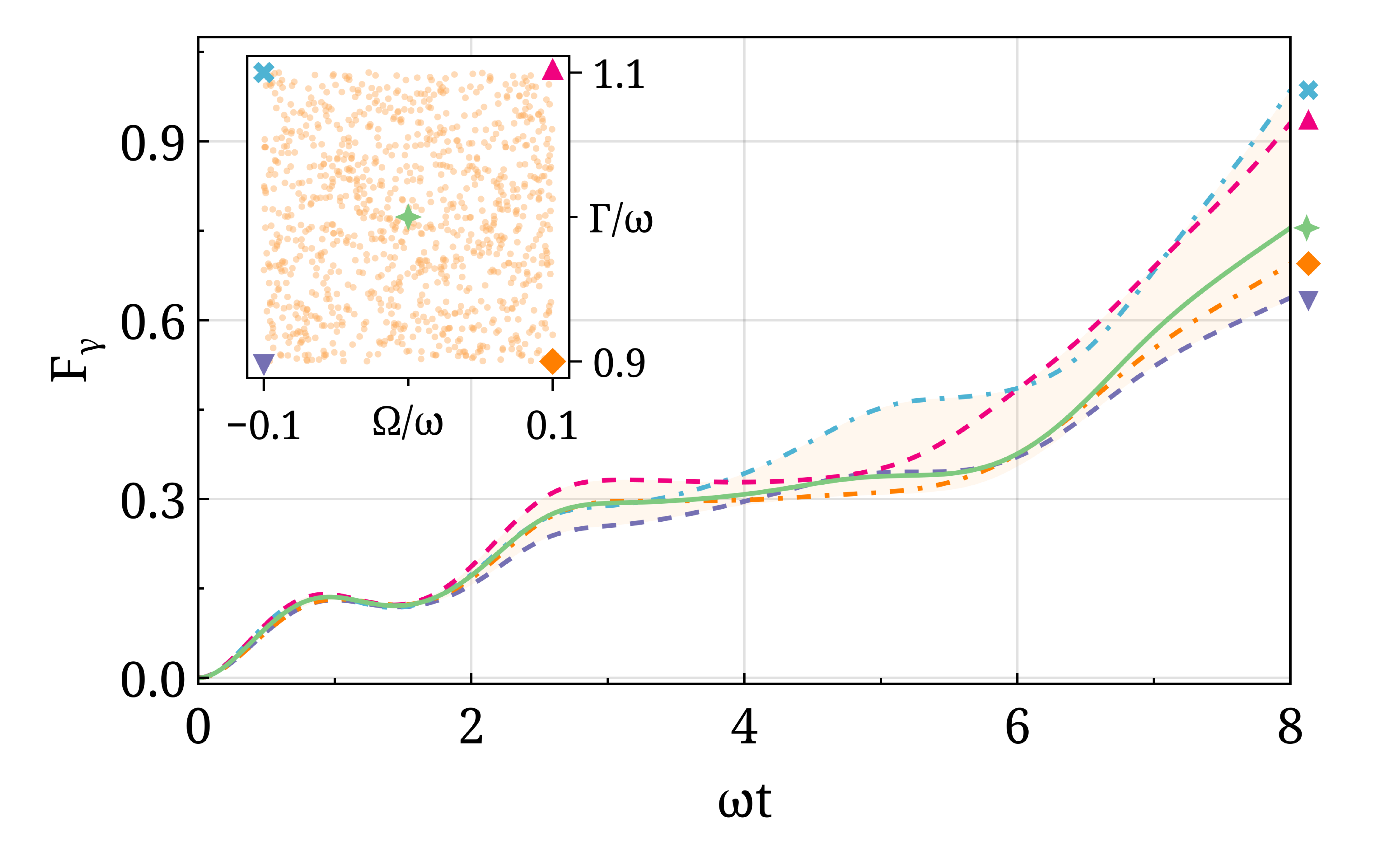}
    \caption{(Color online) Robustness of the estimation to errors in the assumption of other parameters. $1000$ random points plus the 4 corners are taken in a box bound around $\pm 0.1$ of the other two parameters $\Gamma_0 = 1$, $\Omega_0 = 0$ (inset). The ranges of the Fisher information is shown as a function of time in the orange shaded area, the reference true dynamics is shown as the green solid line, and the 4 corner cases are shown as dash(-dotted) lines.}\label{fig_gxerr}
\end{figure}

Note that the whole open system dynamics is determined by $3$ non-Markovian bath spectrum parameters $\{\gamma, \Gamma, \Omega\}$. Here we are focusing on single-parameter estimations, so one needs to assume a value for the other $2$ parameters - here in this example, $\Gamma_0=\omega$ and $\Omega_0=0$. In real-world scenarios, there could be some errors associated with the assumptions of the other system parameters, and we now study how robust this protocol can be against these errors. In Fig.~\ref{fig_gxerr} we chose $1000$ random points around $\pm 0.1\omega$ of the true values of $\Gamma_0$ and $\Omega_0$ for the estimation of the memory parameter $\gamma$. It can be seen that the control is still effective in driving QFI to a peak value at the end of the runtime without too large deviations.

\begin{figure}[t]
    \centering
    \includegraphics[width=.42\textwidth]{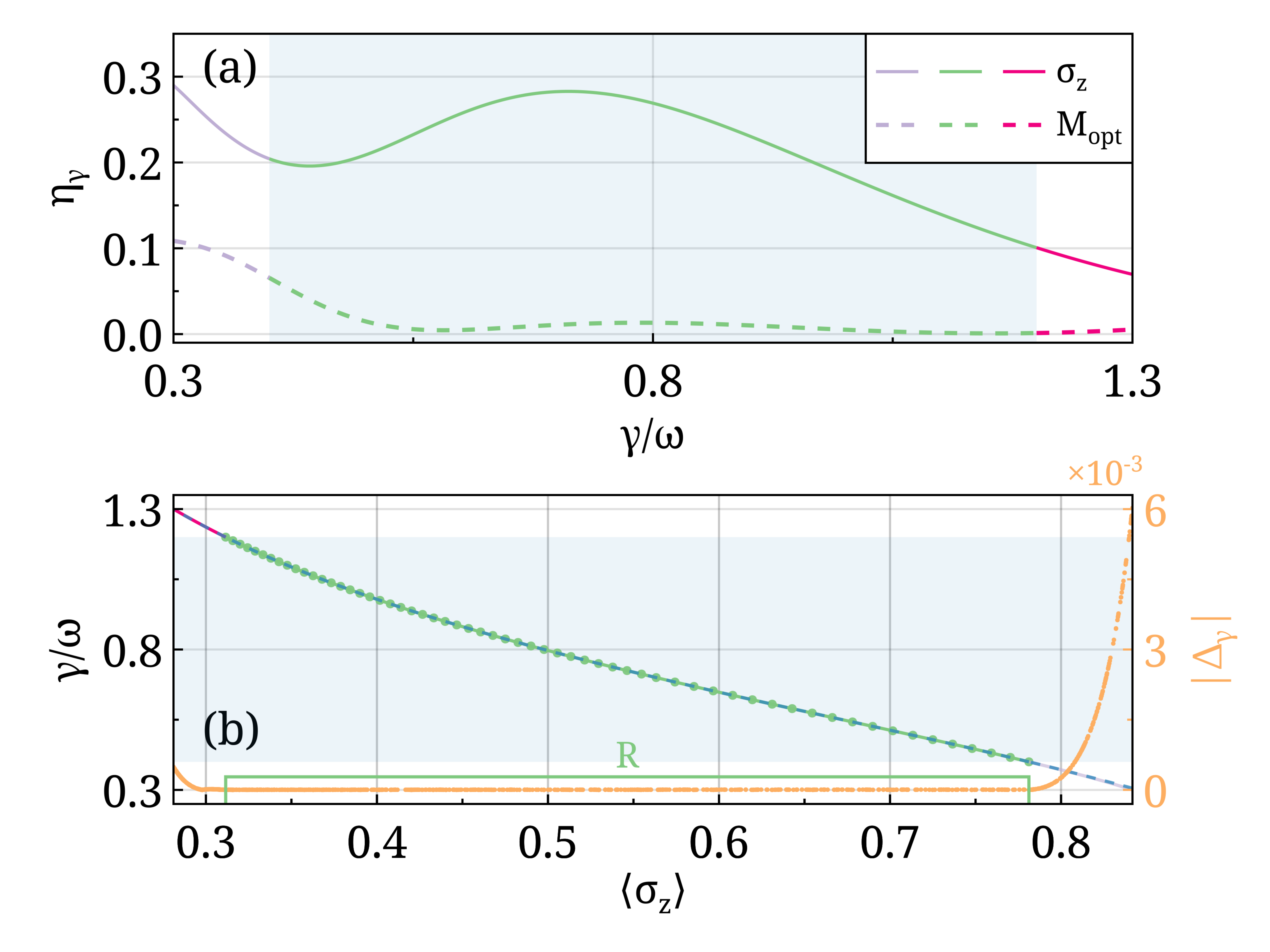}
    \caption{(Color online) Panel (a) The closeness of the uncertainty to the \crb{} (CRB) as $\eta_\gamma = \delta_{A, \theta} \sqrt{F_\theta} - 1$ (value of $0$ indicates saturation of the CRB). Blue shaded area represents the training range R, and the line color scheme is the same as Fig.~\ref{fig_sz_gamma}. The solid lines are from the measurement of $\sigma_z$, while the dashed line is from an optimized measurement on $M_{\rm opt}$. It can be seen that the measurement on $\sigma_z$ can drive the uncertainty quite close to the CRB, whereas the optimized measurement can further bring it close to saturation. Panel (b) Left axis: The expectation value $\langle \sigma_z(T_f) \rangle$ (solid lines) is a monotone of $\gamma$, so the values of $\gamma$ (solid lines) may be read out from this measurement via curve fitting (dashed line) from the data of the training points (green dots), and the green bracket highlights the training range $R$. A good agreement between the two can be observed. Right axis: The discrepancy between the actual value of $\gamma$ and the curve fitting result are depicted as orange dots.} \label{fig_szgamma_unclfit}
\end{figure}

The QFI only gives a lower bound for the uncertainty in the parameter estimation through the \crb{} (CRB)~\cite{Cramer1999a, Cramer1946a} as $\delta_{A,\theta} \geq 1/\sqrt{F_\theta}$ in a single measurement for the estimation of a parameter $\theta$, with uncertainty $\delta_{A,\theta} = \Delta A/|d \langle A \rangle / d \theta|$ for an observable $A$, and variance $\Delta A = \sqrt{\langle A^2 \rangle - \langle A \rangle^2}$. To saturate or approach the CRB, one still needs to find an appropriate physical quantity to measure. As a measure of how close the uncertainty is to the CRB, we rewrite the CRB as $\eta_\theta = \delta_{A, \theta} \sqrt{F_\theta} -1 \geq 0$, where $\eta_\theta = 0$ would indicate saturation of the CRB. Here, we chose $A=\sigma_z$ and plot $\eta_\gamma$ as a function of $\gamma$ as solid lines in Fig.~\ref{fig_szgamma_unclfit} (a), with the same color scheme as Fig.~\ref{fig_sz_gamma}.  It can be observed that the resulting uncertainty is quite close to the CRB. In situations where a near saturation of the CRB is desired, we can optimize the measurement observable as $A=M_{\rm opt}$ to minimize the uncertainty $\delta_{A, \gamma}$. Parameterize $M_{\rm opt} = \sin(\varphi_1)\cos(\varphi_2) \sigma_x + \sin(\varphi_1)\sin(\varphi_2) \sigma_x + \cos(\varphi_1) \sigma_z$, we can optimize for the parameters $\varphi_{1, 2}$ to minimize $\delta_{A, \gamma}$ (see~\cite{supp}). For the parameters considered here, it is found to be $M_{\rm opt}\approx -0.14 \sigma_x -0.4 \sigma_y + 0.91 \sigma_z$. The resulting closeness to the CRB is plotted as dashed lines in Fig.~\ref{fig_szgamma_unclfit} (a). It can be seen that the uncertainty is brought much closer to the CRB in this case to around $1\%$ on average in $R$.
Moreover, in Fig.~\ref{fig_szgamma_unclfit} (b) we plot the estimated $\gamma$ as a function of $\langle \sigma_z \rangle$. Crucially, the $\gamma$ is monotonic against $\langle \sigma_z \rangle$, such that $\langle \sigma_z \rangle$ as a function of $\gamma$ is invertible, and one can directly read out the value of $\gamma$ via a simple curve fitting of $\langle \sigma_z \rangle$ against $\gamma$. In Fig.~\ref{fig_szgamma_unclfit} (b) we show the line-fit against the actual values of $\gamma$, where the line-fit is obtained from the $60$ training points. It can be seen that the curve fitting error is negligible with a maximum error around $6\times 10^{-3}$. In addition, the optimized measurement $M_{\rm opt}$ is also found to be monotonic and invertible~\cite{supp}.

\section{Conclusion}

We employ Krotov optimal control, combined with machine-learning-inspired training, to estimate non-Markovian bath parameters as a dynamical process. The protocol addresses the challenge of determining an optimal measurement time with unknown parameters by training a control field to maximize the QFI at a prescribed runtime. Using numerical automatic differentiation, we optimize QFI- and CRB-based targets even when analytical derivatives are intractable.

Using a two-level-system (TLS) model, we demonstrate that the protocol achieves maximal QFI at the desired runtime and outperforms uncontrolled dynamics. The QFI is enhanced for stronger memory effects (smaller $\gamma$), indicating that quantum memory effects can improve estimation precision. The protocol is robust to uncertainties in both the assumed parameter range and other system parameters.

The resulting measurements approach the CRB and remain monotonic in the parameter of interest, enabling direct extraction via simple curve fitting. The method extends naturally to other bath parameters, such as $\Gamma$ and $\Omega$, and shows similar performance for the TLS decay model~\cite{supp}. More broadly, it applies to a wide class of open quantum systems where sensing is dynamical and an optimal measurement time exists.

\begin{acknowledgments}
 This project is supported by the U.S. Department of Defense (ACC-New Jersey under Contract No. W15QKN-24-C-0004).
\end{acknowledgments}

%%%%%%%%%%%%%%%%%%%%%%%%%%%%%%%%%%%%%

%\bibliographystyle{prs}
% \bibliography{learnou}


%%%%%%%%%%%%%%%%%%%%%%%%%%%%%%%%%%%%%

\clearpage
\onecolumngrid
\clearpage
\pagenumbering{gobble}



\section*{Supplementary material}

# Learning Non-Markovian Noise via Ensemble Optimal Control: Supplementary Materials

Da-Wei Luo[1, *] and Ting Yu[1, †]

[1]*Center for Quantum Science and Engineering and Department of Physics,
Stevens Institute of Technology, Hoboken, New Jersey 07030, USA*

(Dated: April 17, 2026)

Here we present some supporting materials for the main text, including more details about the optimization process, and results for the estimations of other non-Markovian spectrum parameter and TLS dissipative systems.

## I. ON THE ENSEMBLE AVERAGE OF QSD MASTER EQUATIONS, AND OPTIMIZATION OF QFI

For the vectorization of the non-Markovian master equation to cast the problem as Eq. (3) in the main text, we column-stack the density matrix, and uses the Kronecker product property to write, for an operator $O$, the super-operation $\mathcal{S}[O] = AOB$ as

$$
\begin{aligned}
|\mathcal{S}[O]\rangle\rangle &= [AOB]_{mn}|n\rangle|m\rangle \\
&= A_{mi}O_{ij}B_{jn}|n\rangle|m\rangle \\
&= O_{ij}(B_{jn}|n\rangle) \otimes (A_{mi}|m\rangle) \\
&= O_{ij}[B^T|j\rangle] \otimes [A|i\rangle] \\
&= [B^T \otimes A]O_{ij}|j\rangle|i\rangle \\
&= [B^T \otimes A]|O\rangle\rangle
\end{aligned}
\qquad (1)
$$

Since the QFI is dependent on not only the quantum state $\rho_\theta$ but also $\partial_\theta \rho_\theta$, where $\theta$ is the parameter to be estimated, we need to control both at the same time, as a direct sum of the vectorized master equations with some finite difference $dx$ (in the plots taken as $10^{-4}$), $M = \mathcal{L}_0(\theta) \oplus \mathcal{L}_0(\theta+dx) + c(t)\mathcal{H}_c \otimes \mathcal{H}_c$:

$$
\partial_t \begin{bmatrix} |\rho_\theta\rangle\rangle \\ |\rho_{\theta+dx}\rangle\rangle \end{bmatrix} = \left( \begin{bmatrix} \mathcal{L}_0(\theta) & \mathbf{0} \\ \mathbf{0} & \mathcal{L}_0(\theta+dx) \end{bmatrix} + c(t) \begin{bmatrix} \mathcal{H}_c & \mathbf{0} \\ \mathbf{0} & \mathcal{H}_c \end{bmatrix} \right) \begin{bmatrix} |\rho_\theta\rangle\rangle \\ |\rho_{\theta+dx}\rangle\rangle \end{bmatrix}
\qquad (2)
$$

The QFI may then be obtained through [1], with the density operator in an eigen-decomposition $\rho = \sum_i \lambda_i |\lambda_i\rangle\langle\lambda_i|$,

$$
F_\theta = \sum_{\lambda_a+\lambda_b \neq 0} 2\frac{|\langle\lambda_a|\partial_\theta \rho|\lambda_b\rangle|^2}{\lambda_a + \lambda_b}
\qquad (3)
$$

$$
\approx \sum_{\lambda_a+\lambda_b \neq 0} 2\frac{|\langle\lambda_a|(\rho_{\theta+dx} - \rho_\theta)/dx|\lambda_b\rangle|^2}{\lambda_a + \lambda_b}
\qquad (4)
$$

In particular, since here we are considering TLS, it may be simplified as [1]

$$
F_\theta = (\partial_\theta \vec{r}) \cdot (\partial_\theta \vec{r}) + \frac{(r \cdot \partial_\theta \vec{r})^2}{1 - |\vec{r}|^2}
\qquad (5)
$$

where $\vec{r}$ is the Bloch sphere coordinates $\vec{r} = \langle \sigma_{x,y,z} \rangle$, and $\partial_\theta \vec{r} = \text{tr}[(\partial_\theta \rho_\theta)\sigma_{x,y,z}]$ where we can approximate $\partial_\theta \rho_\theta$ with the finite difference.

---

[*] dawei.luo@stevens.edu

[†] Ting.Yu@stevens.edu

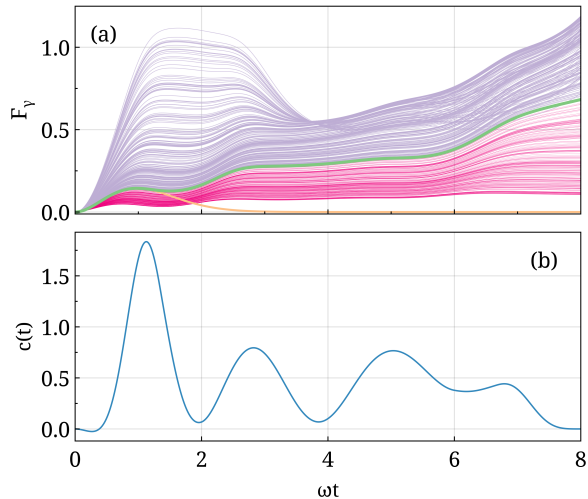

FIG. 1. Panel(a) QFI dynamics for the optimization of final QFI for a single $\gamma_M = 0.8$ (green line), where $\gamma = \gamma_M \pm 0.5$ are shown as pink (purple) lines, and the orange line represents the uncontrolled case. Panel (b) control function as a function of time, $H_c$ taken to be the same as the main text.

Due to the complexity of the QFI expression, finding an analytical expression of $\partial F/\partial s$, where $s = |\rho_\theta\rangle\rangle \oplus |\rho_{\theta+dx}\rangle\rangle$, can be challenging. To solve this issue, we use automatic-differentiation (AD) to calculate the costate $|\chi(T_f)\rangle$, using the Wirtinger notation [2],

$$|\chi(T_f)\rangle = -\frac{\partial J_T}{\partial \langle\psi(T_f)|} \tag{6}$$

$$k\text{-th element: } \chi_k(T_f) = -\frac{\partial J_T}{\partial \psi_k^*(T_f)} \tag{7}$$

$$\Rightarrow |\chi(T_f)\rangle = -\frac{1}{2}\nabla_\psi^{\text{AD}} J_T \tag{8}$$

where $\nabla^{\text{AD}}$ is the numerical AD, here calculated with the `Zygote.jl` library [3, 4].

The ensemble optimization may then be carried out, choosing $n$ points in some training range $\gamma_i \in R$,

$$M_0^{(i)} = \begin{bmatrix} \mathcal{L}_0(\gamma_i) & \mathbf{0} \\ \mathbf{0} & \mathcal{L}_0(\gamma_i + dx) \end{bmatrix} \tag{9}$$

where $\mathcal{L}_0$ is the vectorized form of the QSD master equation, Eq. (7) in the main text, where the dependence on the bath parameters $\{\gamma, \Gamma, \Omega\}$ comes from the $\bar{O}$ operator. It is worth pointing out that since the ensemble optimization control would be effective for, e.g. the midpoint in the range $R$, the reverse may also sometimes be true depending on system parameters or Krotov control initial guess, where the control that target a single point in $R$, such as the mid-point, may also be quite robust for other parameters in $R$. However, the ensemble learning optimization would still be a more reliable and systematic way to find a control that is targeted to optimize the average $J_T$.

In Fig. 1 we show the result where a single $\gamma_M = 0.8\omega$ is considered for the optimization. Comparison with the ensemble optimization are shown in Fig. 2(b) in the main text. It can be seen while the QFI is still driven to some optimal value at $t = T_f$, the achieved QFI is lower than the ensemble optimization results.

## II. NOTE ON THE PULSE UPDATE IN THE QSD NON-MARKOVIAN MASTER EQUATIONS

In addition to needing to re-calculate the effective Lindbladian each iteration after the control functions have been updated, note that the Krotov control's pulse update equation is given by [5–8]

$$\Delta c_l^{(i)}(t) = \frac{S_l(t)}{\lambda_{a,l}}\text{Im}\left[\sum \left\langle \chi_k^{(i-1)} \middle| \left(\frac{\partial H^{(i)}}{\partial c_l^{(i)}(t)}\right) \middle| \psi_k^{(i)} \right\rangle\right] \tag{10}$$

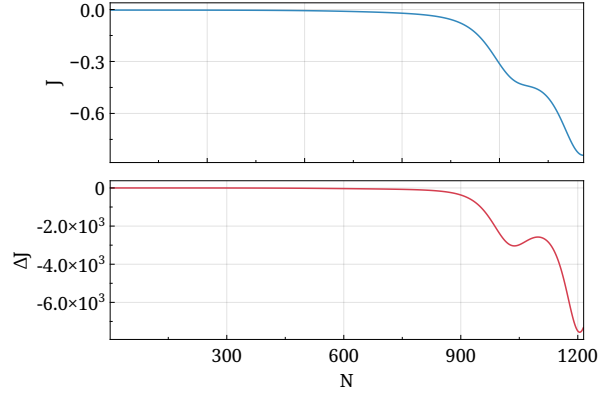

FIG. 2. Monotonic convergence of the algorithm: for the example in the main text on the estimation of $\gamma$, around 1600 iterations are run, with $\lambda = 150$ and $dt = 0.001$.

for the $i$-th iteration, $l$-th control and states labelled by $k$, and $S_l$ is the update envelope shape (usually taken to be the Blackman shape), $\lambda_{a,l}$ is a step size, $H$ is the total effective Hamiltonian in a vectorized differential equation $\partial_t |\psi(t)\rangle = -iH|\psi(t)\rangle$, and $|\chi\rangle$ is the backwards propagated costate. It requires the calculation of

$$\frac{\partial}{\partial c_i(t)} H_{\text{tot}}, \text{ or, here as } \frac{\partial}{\partial c_i(t)} \mathcal{L} \tag{11}$$

While for simple closed system or Markov systems, we only have $H_{\text{tot}} = H_0 + \sum_i c_i(t) H_i$, so the derivative is just $H_i$, in the non-Markovian case the dissipation part of the master equation is given by the $\bar{O}$ operator, which has dependence on the controls $c_i(s)$. However, as a leading order approximation, the term $\partial \bar{O}/\partial c_i(t)$ may be dropped. Consider post-Markov perturbative expression, for $H_s(t) = H_0 + \sum_i c_i(t) H_i$, as [9]

$$\bar{O}(t) = \int_0^t \alpha(t,s) O(t,s,z^*) ds \tag{12}$$

$$= \int_0^t \alpha(t,s) ds \left[ L - i [H_s(s), L] (t-s) - \left( \int_0^s \alpha(s,u) du \, [L^\dagger, L] L \right)(t-s) \right] \tag{13}$$

$$\tag{14}$$

only the second term inside the integral is $c_i(t)$ dependent, but note

$$\frac{\partial}{\partial c_i(t)} \int_0^t \alpha(t,s)(t-s) ds [c_i(s) H_i(s), L] \tag{15}$$

$$= \int_0^t \alpha(t,s)(t-s) \delta(t,s) ds [H_i(s), L] = 0 \tag{16}$$

due to the $(t-s)$ term.

More generally, the noise-free leading order of $\bar{O}$ follows the *non-linear* differential equation

$$\partial_t \bar{O} = \alpha(0) L - \gamma_{\text{eff}} \bar{O} + \left[ -iH - L^\dagger \bar{O}, \bar{O} \right], \tag{17}$$

formally,

$$\bar{O}(t) \approx \bar{O}(t - dt) + dt \left\{ \alpha(0) L - \gamma_{\text{eff}} \bar{O}(t - dt) + \left[ -iH_s(t - dt) - L^\dagger \bar{O}(t - dt), \bar{O}(t - dt) \right] \right\} \tag{18}$$

which leads to, since RHS is at $t - dt$,

$$\frac{\partial}{\partial c_i(t)} \bar{O}(t) \approx 0. \tag{19}$$

Thus, with small enough $dt$ (which the Krotov algorithm already requires), one may approximate

$$\frac{\partial}{\partial c_i(t)} \mathcal{L}[\rho] \approx -i[H_i, \rho]. \tag{20}$$

In Fig. 2 we show the target function $J$ and its change $\Delta J$ for the example considered in the main text. We can see that $J$ monotonically decreases, and $\Delta J$ is negative, indicating that the iterations are monotonically convergent. Note that here, a typical running cost for Eq. (4) in the main text is taken as

$$g_a = \frac{\lambda_{a,l}}{S_l(t)} \left[ c_l^{(i)}(t) - c_{l,\text{ref}}^{(i)}(t) \right]^2 \quad (21)$$

where the reference field $c_{l,\text{ref}}^{(i)}(t)$ is taken to be the one from the previous iteration as a guess. This also means $\Delta J$ needs to be checked against the same reference control fields, in general $\Delta J \neq J^{(i+1)} - J^{(i)}$ [8] but should be calculated as

$$\Delta J = \Delta J_T + \sum_l \int \frac{\lambda_{a,l}}{S_l(t)} \left[ c_l^{(i)}(t) - c_l^{(i-1)}(t) \right]^2 dt \quad (22)$$

for the $i$-th iteration.

## III. OPTIMIZATION OF THE MEASUREMENT OBSERVABLE

Taking $n = 60$ evenly spaced points in $\gamma_i \in [0.3, 1.3]$, we can optimize the measurement observable $M_{\text{opt}} = \sin(\varphi_1)\cos(\varphi_2)\sigma_x + \sin(\varphi_1)\sin(\varphi_2)\sigma_x + \cos(\varphi_1)\sigma_z$, by minimizing the norm of the vector

$$\delta_{\mathbf{M_{opt}}} = [\delta_M(\gamma_1), \delta_M(\gamma_2), \ldots, \delta_M(\gamma_n)], \quad (23)$$

where $\delta_M(\gamma)$ is the uncertainty wrt. $\gamma$, for a measurement of $M_{\text{opt}}$. A blackbox optimization [10] can then solve for the parameters $\varphi_{1,2}$ to minimize the norm. The resulting observable can bring the uncertainty close to the CRB, while also being monotonic so it's reversible so that the parameter $\gamma$ can be obtained through a simple line fit. We show the linefit result in Fig. 3.

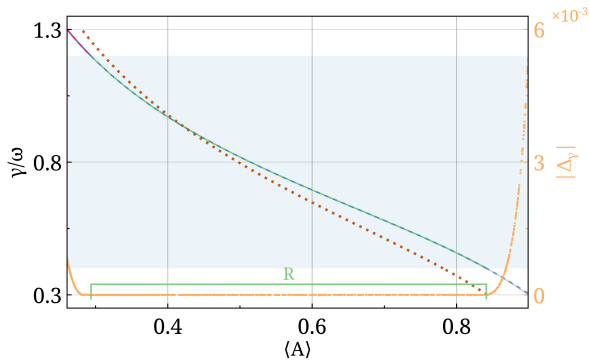

FIG. 3. Line fit result for a measurement of $A = M_{\text{opt}}$ and $A = \sigma_z$ for comparison (left axis). Color schemes are the same as Fig. 4 in the main text. Solid lines and dashed lines are for the optimized measurement (actual value and line-fit value, respectively), and the brown dotted line is for $A = \sigma_z$ as comparison. The errors between the actual values and the line-fit values are plotted as orange dots (right axis).

In the following sections, we will present the results for the estimation of other bath spectrum parameter such as central frequency $\Omega$ and coupling rate $\Gamma$. The dissipative model $L = \sigma_-$ is also studied. The same color scheme as the main text is used throughout (i.e. green for data in the training range, pink for above training range and purple for below training range).

## IV. ADDITIONAL PLOTS FOR DEPHASING MODEL, ESTIMATION OF THE CENTRAL FREQUENCY $\Omega$

In addition to the estimation of the memory parameter $\gamma$, we can also use this protocol for the other parameters.

For example, the results for the estimation of the central

frequency $\Omega$ is displayed in Fig. 5. Here, the total runtime is set to $T_f = 12$, and 10 training values are taken for $\Omega \in R = [0, 9, 1.1]$, with $\Gamma = 1$ and $\gamma = 0.7$. In addition, we also considered $\Omega$ in $\pm 0.1$ of the training range $R$ for a test of robustness. 1000 random test points are chosen, with 500 in $R$ and 250 above (below) $R$. In this case, we can see that the control is highly effective and QFI may be greatly enhanced from its uncontrolled state. All color schemes are the same as the main text (green, magenta, and purple lines refers to within, above and below training regions).

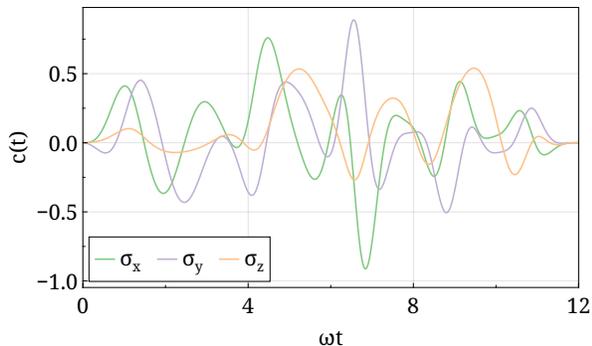

FIG. 4. (Color online) Estimation of $\Omega$ of the dephasing model, control field.

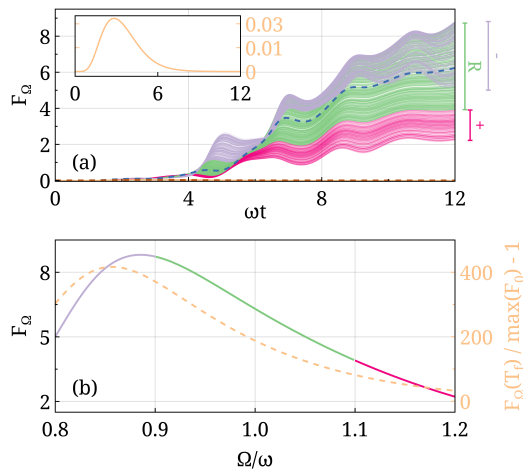

FIG. 5. (Color online) Estimation of the central frequency $\Omega$. Panel (a) Fisher information as a function of time, line colors follow the same scheme as in Fig. 2 in the main text. Note here the achieved Fisher information of the uncontrolled case is much less than the controlled case (inset shows a typical case of $\Omega = 1$). Panel (b): final Fisher information as a function of $\Omega$ (left axis, solid lines) and the ratio between the final Fisher information $F(T_f)$ of the controlled and the peak Fisher information of uncontrolled cases $\max(F_0)$ as $F(T_f)/\max(F_0) - 1$ (right axis, dashed lines). This ratio is found to be larger than 1, indicating that the QFI is larger in the controlled case.

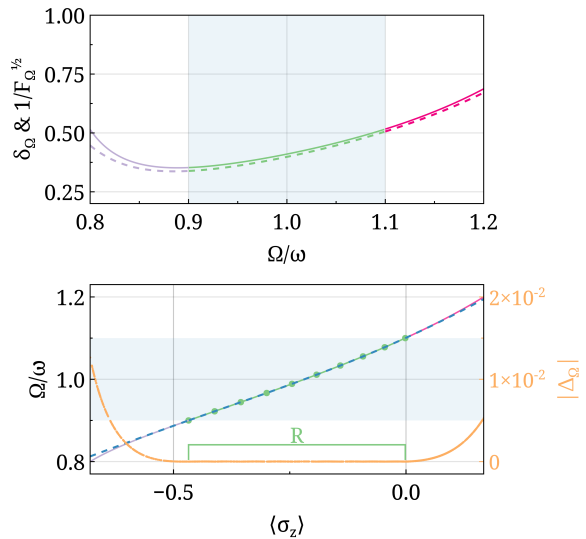

FIG. 6. (Color online) Estimation of $\Omega$ of the dephasing model, uncertainty and line fit. Measurement is on $\sigma_z$.

## V. ADDITIONAL PLOTS FOR DEPHASING MODEL, ESTIMATION OF THE SYSTEM-BATH COUPLING $\Gamma$

Here we consider the estimation of $\Gamma$, with runtime $T_f = 8$, $\gamma = 0.7$ and $\Omega = 0$. 10 training points are taken for $\Gamma \in R = [0.5, 0.8]$. For robustness test, $\Gamma$ in $\pm 0.1$ of the training range is also being considered, with 1000 random tests in $R$ and 300 random points above (below) $R$.

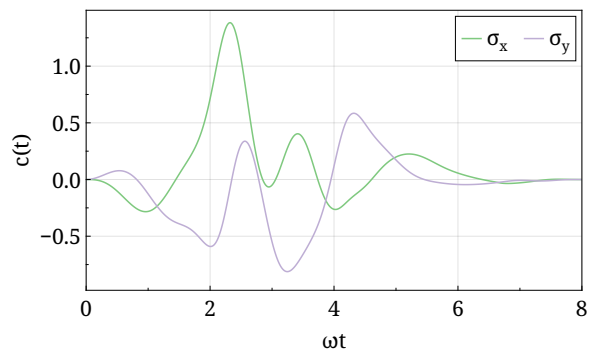

FIG. 7. (Color online) Estimation of the coupling strength $\Gamma$, for the dephasing model: control field.

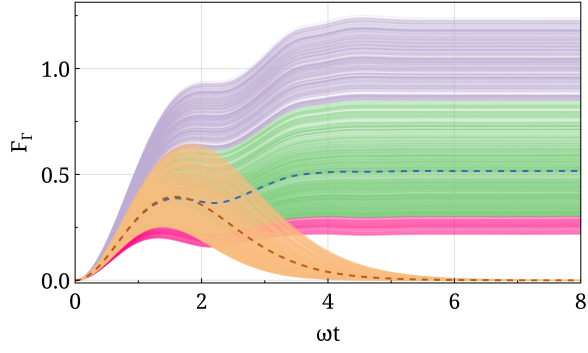

FIG. 8. (Color online) Estimation of the coupling strength $\Gamma$, for the dephasing model: QFI dynamics.

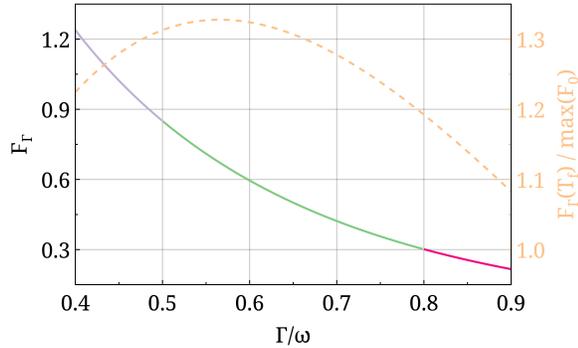

FIG. 9. (Color online) Estimation of the coupling strength $\Gamma$, for the dephasing model: final QFI, and comparison with the uncontrolled cases.

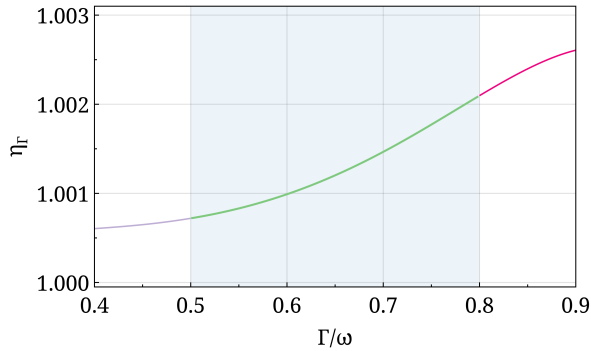

FIG. 10. (Color online) Estimation of the coupling strength $\Gamma$, for the dephasing model: uncertainty and QFI, where the measurement is on $\sigma_z$. Here, since the two curves are too close (indicating a close saturation of the Cramér-Rao bound), we plot $\delta f_\Gamma = \delta_\Gamma \sqrt{F_\Gamma} \geq 1$, where closer to 1 means closer to the Cramér-Rao bound.

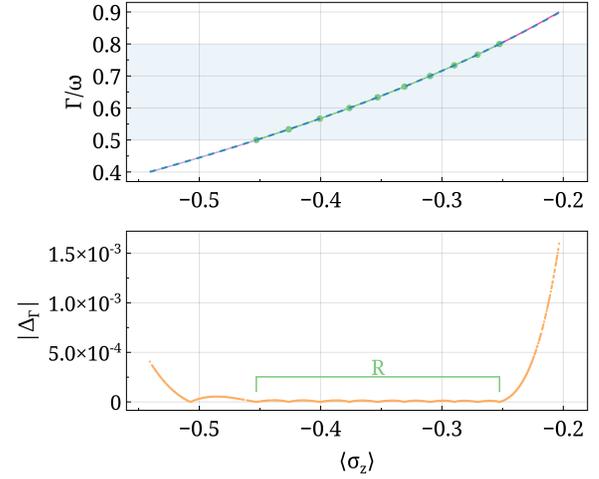

FIG. 11. (Color online) Estimation of the coupling strength $\Gamma$, for the dephasing model: measurement of $\sigma_z$ and curve fitting results.

## VI. ADDITIONAL PLOTS FOR: DECAYING MODEL, ESTIMATION OF THE MEMORY PARAMETER $\gamma$

Here we consider the TLS decaying model, $H_0 = \sigma_z$ and $L = \sigma_-$. Similar results may be obtained for the ensemble learning control for the parameter estimation. We considered $T_f = 90$, $\Gamma = 1.2$ and $\Omega = 0$. 10 training points are taken for $\gamma \in [0.7, 1]$, and values $\pm 0.1$ the training range are considered as tests for robustness, with 1000 random tests in $R$ and 300 random points above (below) $R$.

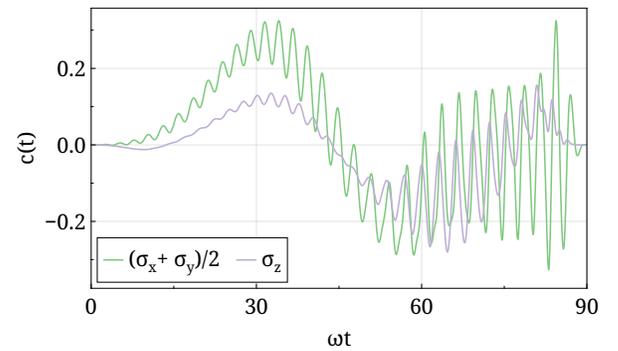

FIG. 12. (Color online) Estimation of the memory parameter $\gamma$ of the decaying model, control fields.

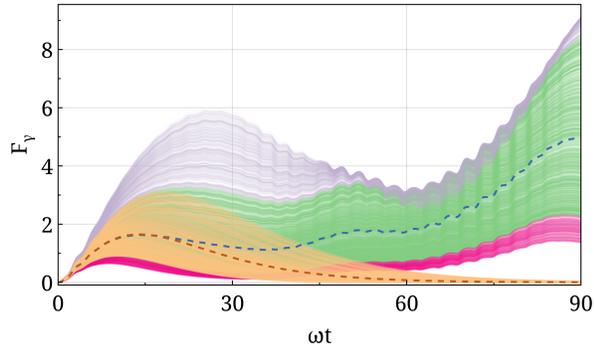

FIG. 13. (Color online) Estimation of the memory parameter $\gamma$ of the decaying model, QFI dynamics.

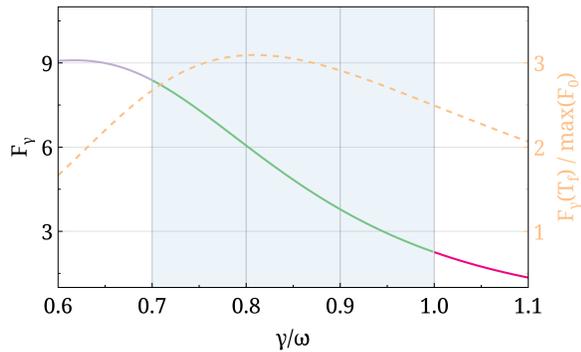

FIG. 14. (Color online) Estimation of the memory parameter $\gamma$ of the decaying model, final QFI and comparison with the uncontrolled case.

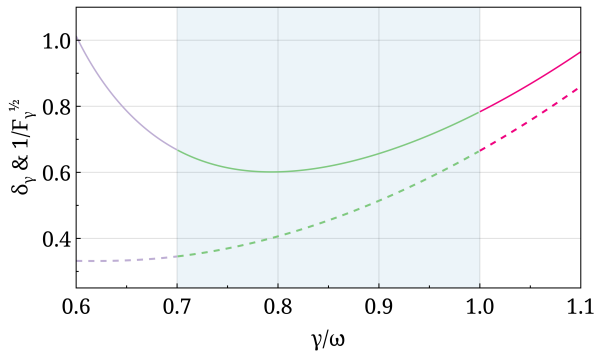

FIG. 15. (Color online) Estimation of the memory parameter $\gamma$ of the decaying model: uncertainty and QFI, measurement is on $\sigma_z$

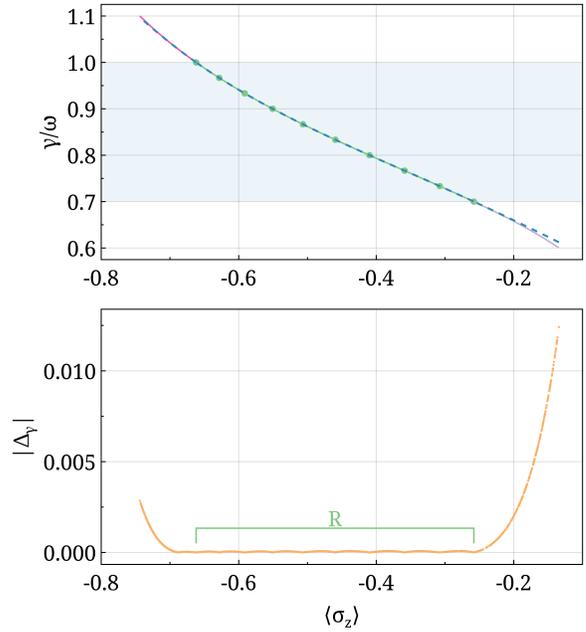

FIG. 16. (Color online) Estimation of the memory parameter $\gamma$ of the decaying model: curve fitting with a measurement on $\sigma_z$.



\end{document}